\documentclass[%
 reprint,
superscriptaddress,
 amsmath,amssymb,
 aps,
prb,
]{revtex4-1}

\usepackage{siunitx}
\usepackage{natbib}
\usepackage{amsmath}
\usepackage{graphicx}
\usepackage{soul,color}
\usepackage{lipsum}
\usepackage{caption}

\begin{document}

\title{Unveiling Symmetry Protection of Bound States in the Continuum with Terahertz Near-field Imaging}

\author{Niels J.J.van Hoof}
\affiliation{%
Institute for Photonic Integration, Department of Applied Physics, Eindhoven University of Technology, P.O. Box 513, 5600 MB Eindhoven, The Netherlands
}

\author{Diego R. Abujetas}
\affiliation{%
 Instituto de Estructura de la Materia (IEM-CSIC), Consejo Superior de Investigaciones Cient\'{\i}ficas,\\ Serrano 121, 28006 Madrid, Spain
}%

\author{Stan E.T. ter Huurne}
\author{Francesco Verdelli}
\author{Giel C.A. Timmermans}
\affiliation{%
Institute for Photonic Integration, Department of Applied Physics, Eindhoven University of Technology, P.O. Box 513, 5600 MB Eindhoven, The Netherlands
}

\author{Jos\'e A. S\'anchez-Gil}
\affiliation{%
 Instituto de Estructura de la Materia (IEM-CSIC), Consejo Superior de Investigaciones Cient\'{\i}ficas,\\ Serrano 121, 28006 Madrid, Spain
}%

\author{Jaime G\'omez Rivas}%
\email{j.gomez.rivas@tue.nl}
\affiliation{%
Institute for Photonic Integration, Department of Applied Physics, Eindhoven University of Technology, P.O. Box 513, 5600 MB Eindhoven, The Netherlands
}




\begin{abstract}
Bound states in the continuum (BICs) represent a new paradigm in photonics due to the full suppression of radiation losses. However, this suppression has also hampered their direct observation. By using a double terahertz (THz) near-field technique that allows the local excitation and detection of the THz amplitude, we are able to map for the first time the electromagnetic field of BICs over extended areas, unveiling the field-symmetry protection that suppresses far-field radiation. This investigation, done for metasurfaces of dimer rod resonators, reveals the in-plane extension and formation of BICs with anti-symmetric phases, in agreement with coupled-dipole calculations. By displacing the rods, we demonstrate that mirror symmetry is not a necessary condition for BIC formation. Only $\pi$-rotation symmetry is required, making BICs exceptionally robust to structural changes. This work makes the local field of BICs experimentally accessible, which is crucial for the engineering of cavities with infinite lifetimes.

\end{abstract}

\maketitle

\section{Introduction} 
One of the major drivers in the fields of optics and photonics is the realization and investigation of systems supporting high-quality (high-Q) electromagnetic resonances. These investigations are motivated by fundamental phenomena, such as light-matter interaction, non-linear optics, ultra-strong coupling, or optical switching, as well as by potential applications, such as optical sensing, storage, low threshold lasing, or frequency and polarisation filtering~\cite{Weimann2013,Kodigala2017a,Doeleman2018,Koshelev2018a,Romano2018,Bahari2019}. The quality of a resonator is inherently limited by the losses in the system, which have two possible sources: intrinsic material losses or the absorption of the electromagnetic (EM) wave, and radiation leakage of the EM energy out of the system. The first source of loss can be suppressed or minimized by using materials with no or weak absorption at the resonant frequency (e.g. dielectrics at optical frequencies and/or noble metals at low frequencies). The suppression of radiation leakage is more subtle as it requires to create a fully closed system, which does not allow either to couple far-field EM energy into the resonances, nor to probe these resonances in the far-field. 

Recent advances with more complex periodic structures of scatterers has led to the realization of the so-called Bound States in the Continuum (BICs), which have boosted the fundamental research on high-Q resonators~\cite{Marinica2008,Hsu2013,Blanchard2016,Hsu2016a,Kodigala2017,Koshelev2018b,Ha2018,Carletti2018,Abujetas2019c,Zito2019,Han2019,Wu2020}. These structures define open systems with eigenstates that have their frequency and wave vector in the radiation continuum (inside the light cone) and, consequently, cannot couple to this continuum. Two different types of BICs can be realized in periodic systems or metasurfaces: Symmetry protected BICs and accidental BICs~\cite{Hsu2016a,Koshelev2018b,Abujetas2019c,Abujetas2019d,Liu2019d,Kupriianov2019,Cong2019,Tan2020,Liang2020,Koshelev2019a,Zhao2020}. A symmetry protected BIC can only be generated in the 0th-order of reflection/transmission or the $\Gamma$ point of the periodic system. The symmetry of the unit cell precludes radiation normal to the plane of the array~\cite{Koshelev2018b}. Accidental BICs are formed when two different modes (i.e., different emission patterns) interfere destructively and cancel each other in the far-field under a specific angle of incidence/scattering. Similar to closed systems, the remarkable suppression of coupling to the continuum makes it impossible to directly excite and observe BICs with far-field spectroscopic techniques~\cite{Monticone2019}. Therefore, research on BICs has been so-far limited to the observation of quasi-BICs or sharp Fano resonances in their evolution towards BICs~\cite{Doeleman2018,Taghizadeh2017,Bulgakov2017b,Abujetas2019d,Liu2019d,Kupriianov2019,Cong2019,Tan2020,Liang2020}. 

In this manuscript, we demonstrate that it is possible to excite symmetry protected BICs and directly image the spatial distribution of their associated EM fields. This demonstration is done at THz frequencies by imaging the electric field distribution in the time domain using a double near-field microscopy technique and a metasurface of gold dimer resonators (array of two equal resonators per unit cell). The simultaneous measurement of amplitude and phase on the metasurface enables us to experimentally unravel the mechanisms leading to the complete suppression of radiation losses and the storage of EM energy into the open cavity defined by the metasurface. By displacing the rods in the dimer, we break the mirror symmetry of the metasurface and show that only $\pi$-rotation symmetry is necessary to sustain symmetry protected BICs. This property makes BICs exceptionally robust and the realization of open optical cavities with infinite life times or Q-factors surprisingly simple. 

The strategy to suppress material losses depends on the frequency range. The high permittivity of noble metals at low frequencies results in a small electronic skin depth and low material losses. Noble metals such as gold are, therefore, ideal to manufacture resonant structures at giga- and terahertz frequencies. The arrangement of these structures in a photonic lattice can tailor their radiation properties to enhance scattering in certain directions. These enhanced differential scattering cross sections by interference of the scattered waves is the mechanism leading to applications, such as antenna arrays and beam shaping ~\cite{Bahari2019, Ha2018}. The recent discovery that arrays of similar resonant dipolar scatterers can also fully suppress all radiation channels and support symmetry protected BICs has opened the possibility of realizing ultra-high Q resonances in extended open cavities of scatterers~\cite{Liu2019d,Koshelev2019a,Bulgakov2014}.

\section{Sample description} 
\begin{figure}[h!]
\centering
\includegraphics[width=0.4\textwidth]{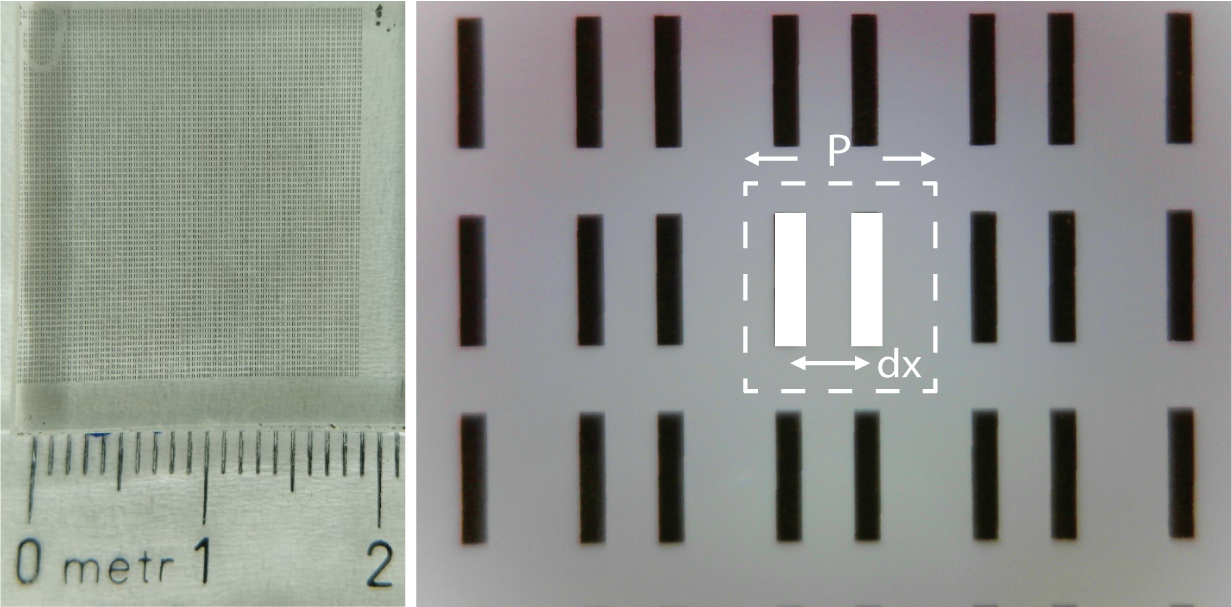}
\caption{(a) Photograph of the sample and (b) microscope image of several unit cells of the sample. The period $P$ and distance between rods $\mathrm{d}x$ are also indicated in the image.}
\label{fig:1_1}
\end{figure}

We have fabricated samples supporting BIC states using standard optical lithography, metal deposition, and lift-off techniques (see supplemental information). A photograph of the sample with dimensions $2 \times 2$ $\mathrm{cm}^{2}$ is shown in Fig.~\ref{fig:1_1}. The investigated sample consists of a 2D periodic lattice of gold rods with a height of 100 nm on top of Z-cut quartz substrate with a thickness of 2.5 mm. A 2 nm thick layer of titanium was used for the adhesion of the gold layer. Figure 1 also shows a microscope image of the sample. The gold rods have a length of 200 $\mathrm{\mu}$m and a width of 40 $\mathrm{\mu}$m, and are designed to have the fundamental ($\lambda/2$) resonance around 0.4 THz. A squared unit cell with periodicity $P=300$ $\mathrm{\mu}$m contains two of these rods, which are displaced from each-other along the diffraction axis by $x= 120$ $\mathrm{\mu}$m. 

\section{Quasi-BIC dispersion calculations} 
\begin{figure*}
\centering
\includegraphics[width=0.95\textwidth]{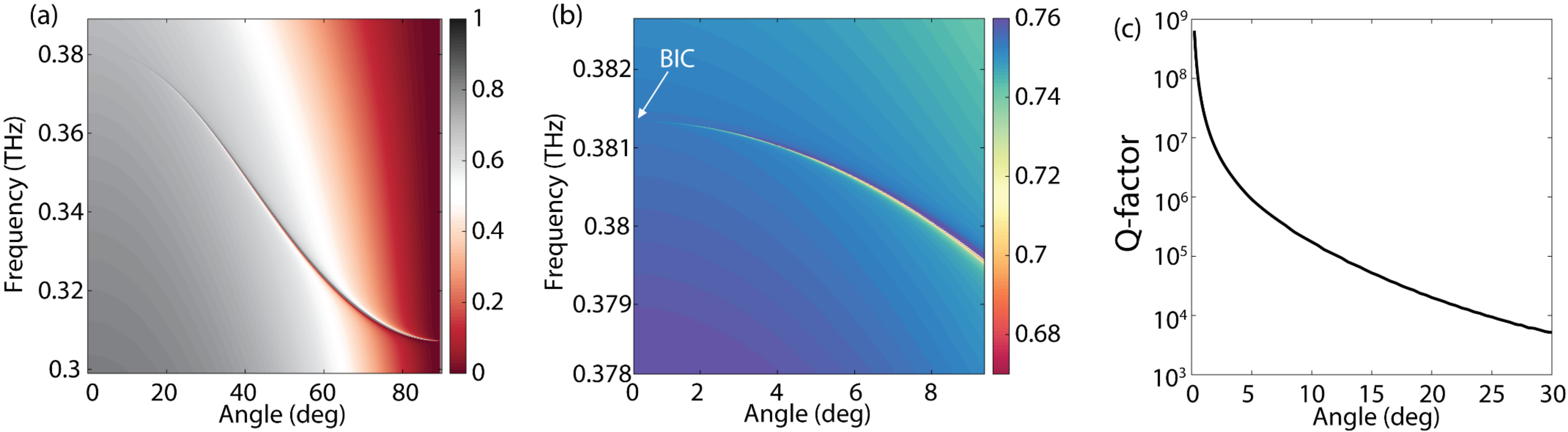}
\caption{(a) Zeroth-order transmittance as a function of the angle of incidence as calculated using the CEMD model. (b) Same as (a) at angles close to normal incidence (note the different scale of the transmittance in (a) and (b). (c) Q-factor as a function of the angle of incidence.}
\label{fig:1_2}
\end{figure*}

To validate the presence of a BIC in the fabricated metasurface, a coupled electric and magnetic dipole (CEMD) calculation is used that fully accounts for all dipole interactions across the infinite array~\cite{Abujetas2018a,Abujetas2020a}. In this particular geometry, each rod is replaced by a longitudinal electric dipole along the polarization of the incident EM wave. The polarizability $\alpha_y$ of the dipole is equal to that of a rod as extracted from a full numerical calculation using the method of moments~\cite{SCUFF1,SCUFF2} (see supplemental information). The expected frequency of the BIC can be extrapolated from the dispersion curve obtained for the zeroth-order transmittance as a function of the angle of incidence and calculated using the CEMD model. This dispersion curve is shown in Fig. \ref{fig:1_2}(a) from 0$^{\circ}$ to 90$^{\circ}$ and at angles close to normal incidence ($\Gamma$ point) in Fig. \ref{fig:1_2}(b). In addition, the angular dependence of the Q-factor is shown in Fig. \ref{fig:1_2}(c), revealing the characteristic divergence of symmetry protected BICs upon approaching normal incidence. We also note the remarkable high value of the Q-factor for all angles despite the very simple geometry, with $Q > 10^4$ for angles of incidence smaller than $25^{\circ}$.

The dimensions of the particles and their periodicity in the sample was chosen such that the radiative coupling between rods in the plane of the array is enhanced by diffraction, which leads to the so-called surface lattice resonances (SLRs)~\cite{Wang2018,Kravets2018}. The presence of two rods per unit cell of the array should lead to two SLRs in which the two rods in the unit cell oscillate in-phase or out-of-phase~\cite{Humphrey2016,Mitrofanov2018,VanHoof2019,Abujetas2019c}. The in-phase or symmetric SLR corresponds to a super-radiant mode that couples very efficiently to the continuum; while the out-of-phase or anti-symmetric SLR defines the BIC with a sub-radiant character. Using the CEMD model, SLRs can be found as the solution of the homogeneous problem (no source), obtained by imposing that the inverse of the scattering matrix vanishes. At normal incidence, the imaginary components of the eigenvalues of these anti-symmetric and symmetric modes calculated using the CEMD model can be approximated by~\cite{Abujetas2019c}
\begin{equation}
Im\left[\Lambda^+\right]=Im\left[\frac{(k^2\Delta \alpha_y)^2}{8G_{yy}}\right] \;,
\label{eq:1}
\end{equation}
\begin{equation}
Im\left[\Lambda^-\right]=Im\left[2\left(\frac{1}{k^2\alpha_y}-G_{byy}\right)-\frac{(k^2\Delta \alpha_y)^2}{8G_{yy}}\right] \;, 
\label{eq:2}
\end{equation}
where $\alpha_y$  is the average polarizability of both particles, $\Delta \alpha_y$ defines the detuning between the two rods in the unit cell, $G_{byy}$ is the self-interaction of each dipole array, and $G_{yy}$ is the mutual interaction of both arrays. In the limit of $\Delta \alpha_y = 0$, i.e.,  when the two rods are equal in size,  $\Lambda^+$ and $\Lambda^-$ correspond to the eigenmodes for out-of-phase and in-phase oscillation of identical dipoles, respectively. Moreover, in the absence of diffraction orders, as it is the case for Eqs.~\ref{eq:1} and~\ref{eq:2}, the imaginary component of the anti-symmetric lattice resonance vanishes. In this case, $\Lambda^+$ can have solutions at real frequencies, which correspond to the symmetry protected BIC. As we demonstrate experimentally below, the out-of-phase oscillation of the two dipoles of equal strength and same frequency is responsible for the complete suppression of the radiation leakage to the far-field.

\section{THz transmission measurements}
\begin{figure}[h!]
\centering
\includegraphics[width=0.4\textwidth]{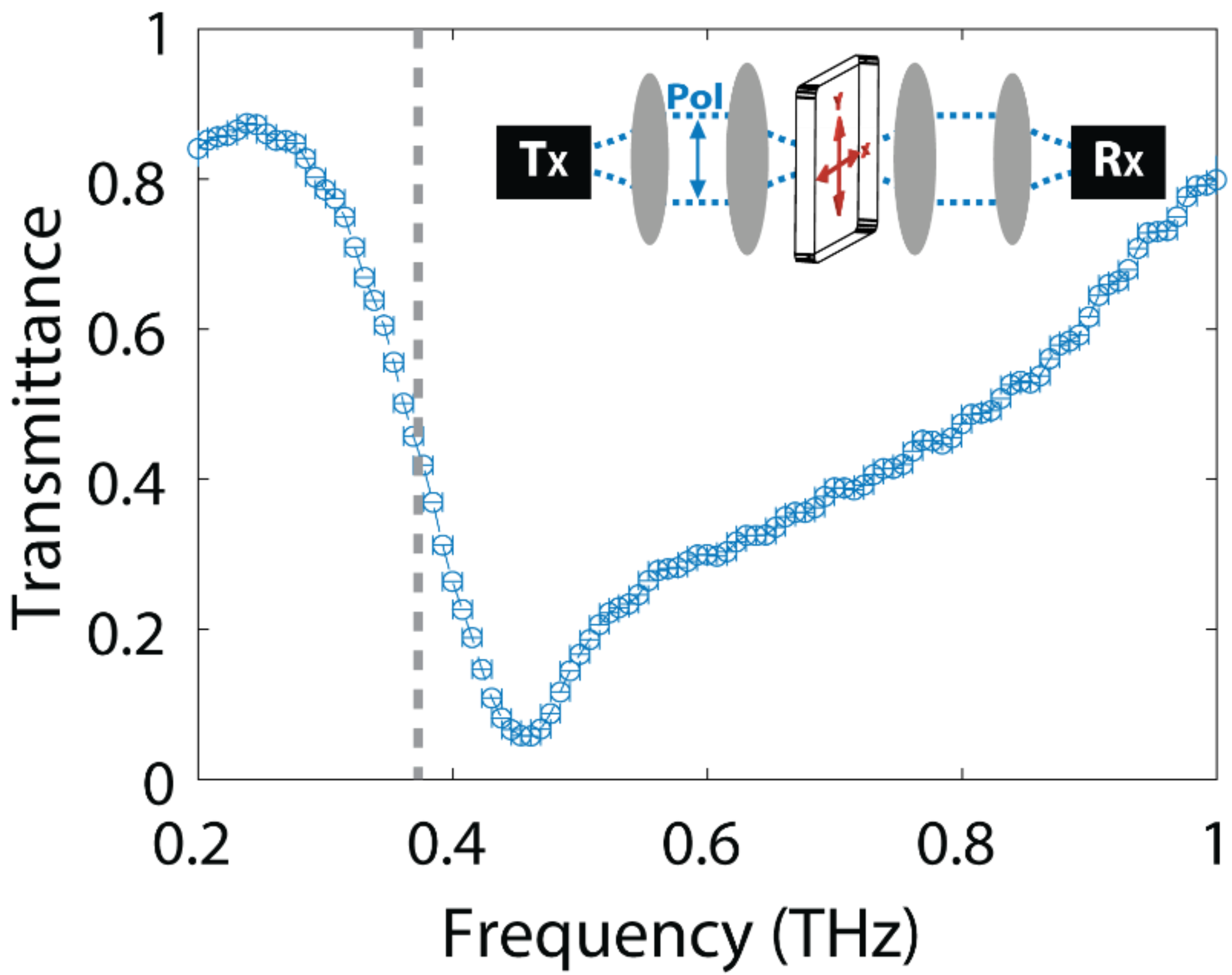}
\caption{Measured far-field transmittance spectrum under normal incidence using a THz time-domain spectrometer, as schematically depicted in the inset.  The expected frequency of the BIC, as obtained from the CEMD model, is indicated by the grey dashed line. The spectral resolution of the measurements is 7 GHz.}
\label{fig:1_3}
\end{figure}

The transmittance through the sample (power transmission normalized by the incident power) was measured with a THz time-domain spectrometer (THz-TDS), as illustrated in the inset of Fig. \ref{fig:1_3}. The transmittance measurement is shown in the same figure, where no resonance associated to the BIC can be observed at 0.38 THz as could be expected from the suppression of coupling to the continuum. Only a very broad resonance at 0.45 THz is visible, which can be associated to the symmetric SLR with increased radiation losses (super-radiance) when the two rods are equal in size, i.e., when $\Delta \alpha_y=0$ and $Im\left[\Lambda^-\right]$ has a maximum value, as given in Eq.~\ref{eq:2}.

We have also done angle dependent transmittance measurements (see supplemental information, Fig.~\ref{fig:S1}). However, the extremely narrow resonance predicted by the CEMD calculations cannot be resolved in these measurements as the frequency resolution of the THz time-domain spectrometer is 7 GHz, as can be seen in the measurement of Fig. \ref{fig:1_3} and limited by reflections in the sample substrate, while the expected resonance linewidth is in the MHz range.

\section{THz near-field measurements of BICs} 

\begin{figure}
\includegraphics[width=0.45\textwidth]{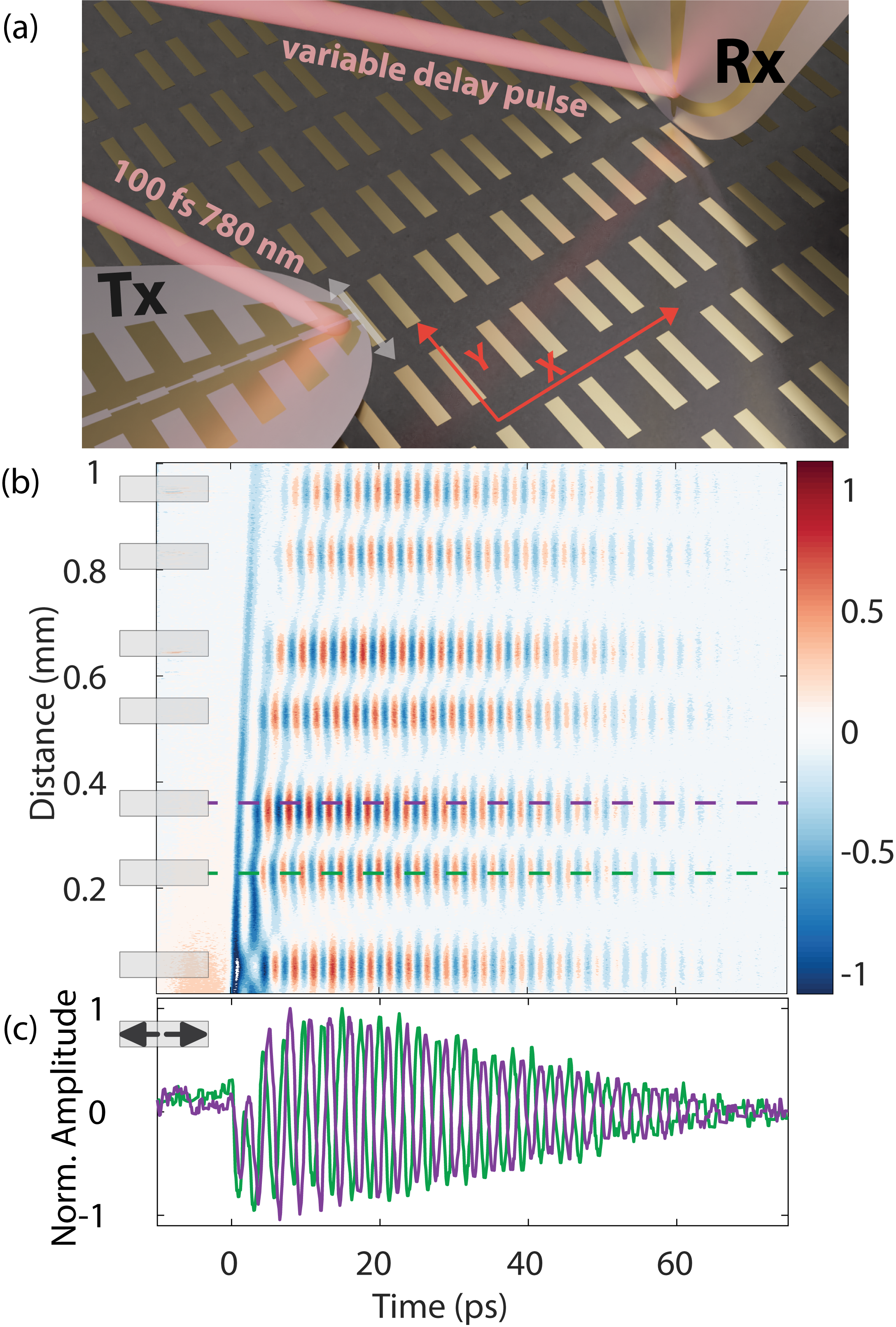}
\caption{(a) Schematic representation of the THz near-field microscope used for the excitation and detection BICs. (b) Normalised THz-field amplitudes as a function of time, plotted at increasing distances from \textit{Tx}, which is located at -60 $\mathrm{\mu}$m as graphically depicted by the black arrows on the lowest rod. Long-lasting oscillations of the electric field and a phase change of 180$^{\circ}$ between rods in a dimer are visible. This out-of-phase behavior is highlighted in (c) by plotting the field amplitudes at two neighboring rods marked by the green and purple dashed lines in (b).}
\label{fig:2_illustration}
\end{figure}

The complete suppression of in- or out-coupling of electromagnetic radiation from the far-field, makes the observation of BICs only possible with near-field techniques~\cite{Sadrieva2019}. As we show next, the field distribution associated with the symmetry protected BICs can be directly imaged by the local excitation of one of the rods in a unit cell of the array and the near-field detection of the THz field on the surface of the 2D array. This local excitation and detection of ultra-short THz pulses is realized by using two separate micro-structured photo-conductive antennas, as depicted in Fig.~\ref{fig:2_illustration}(a). The operating principle of this setup is similar to that of a conventional THz-TDS system. A 100 fs optical pulse from a laser ($\lambda=780$ nm) generates a single-cycle THz pulse in a micro-structured photo-conductive antenna (emitter) \textit{Tx} with a certain dipole orientation depicted by the grey arrow in Fig.~\ref{fig:2_illustration}(a). This THz pulse can be detected with a second micro-structured photo-conductive antenna (receiver) \textit{Rx}, which is gated with another optical pulse that has passed through a computer controlled optical delay line. THz pulses can be detected on ultra-short time scales by varying the arrival time of the second pulse, obtaining both amplitude and phase information. By scanning \textit{Tx} or \textit{Rx}, these transient signals can be mapped in space, imaging the ultra-fast THz field at arbitrary positions on the surface of the array. The small size of the emitter (the gap of the photo-conductive switch is 5 µm) makes it possible to approximate it as a dipole emitter. This approximation is justified with measurements of the radiated THz field by the emitter and its propagation in free space that can be found in the supplemental information (Fig.~\ref{fig:S2}).

The local excitation and near-field detection of the THz field amplitude is done without introducing a significant perturbation to the modes of the sample due to small dimensions of the photo-conductive antennas~\cite{Bhattacharya2016}. Therefore, this technique offers an unique opportunity for the investigation of the excitation and THz-field distribution in resonant structures and metasurfaces. A movie of the THz field amplitude on the 2D periodic lattice of gold rods can be found in the supplemental information. For this measurement, as well as for the results discussed in the manuscript, the THz emitter was positioned in the middle of a gold rod situated in the center of the array and at a height of 3 $\mathrm{\mu}$m above the rod. The measurements started 100 $\mathrm{\mu}$m away to avoid unwanted laser reflections from \textit{Tx} that could damage \textit{Rx} as these photo-conductive antennas are inherently fragile.

\begin{figure}
\centering
\includegraphics[width=0.50\textwidth]{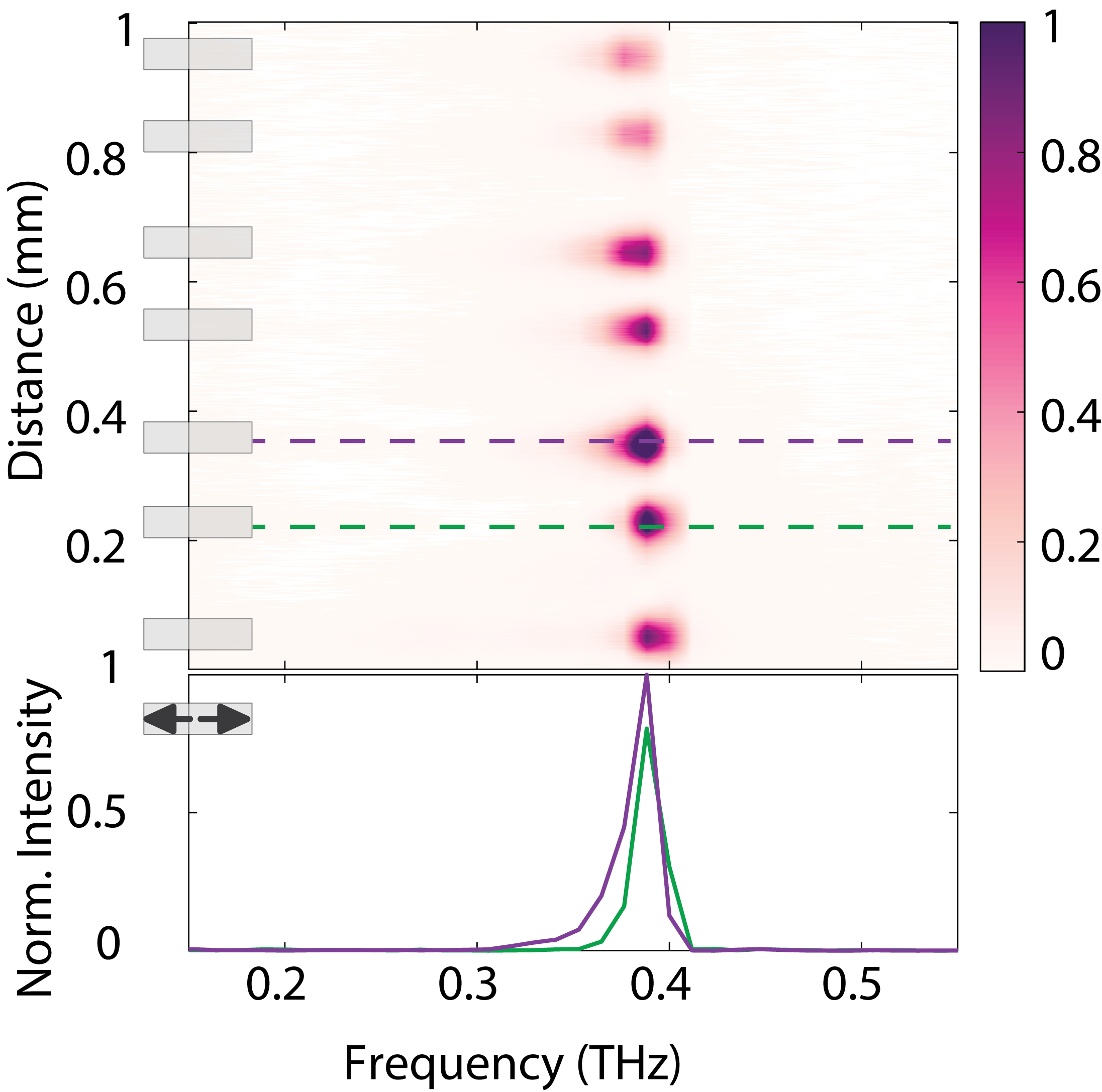}
\caption{Normalised field intensity as a function of frequency at increasing distances from a dipole source located at -60 $\mathrm{\mu}$m, as graphically depicted by the black arrows on the lowest rod on the left. The measured peak frequency agrees with the theoretically predicted BIC at 0.38 THz, as can be better observed from the cross cuts at the positions of the two rods in a dimer shown at the bottom figure}
\label{fig:3_spectra}
\end{figure}

\begin{figure*}
\centering
\includegraphics[width=0.90\textwidth]{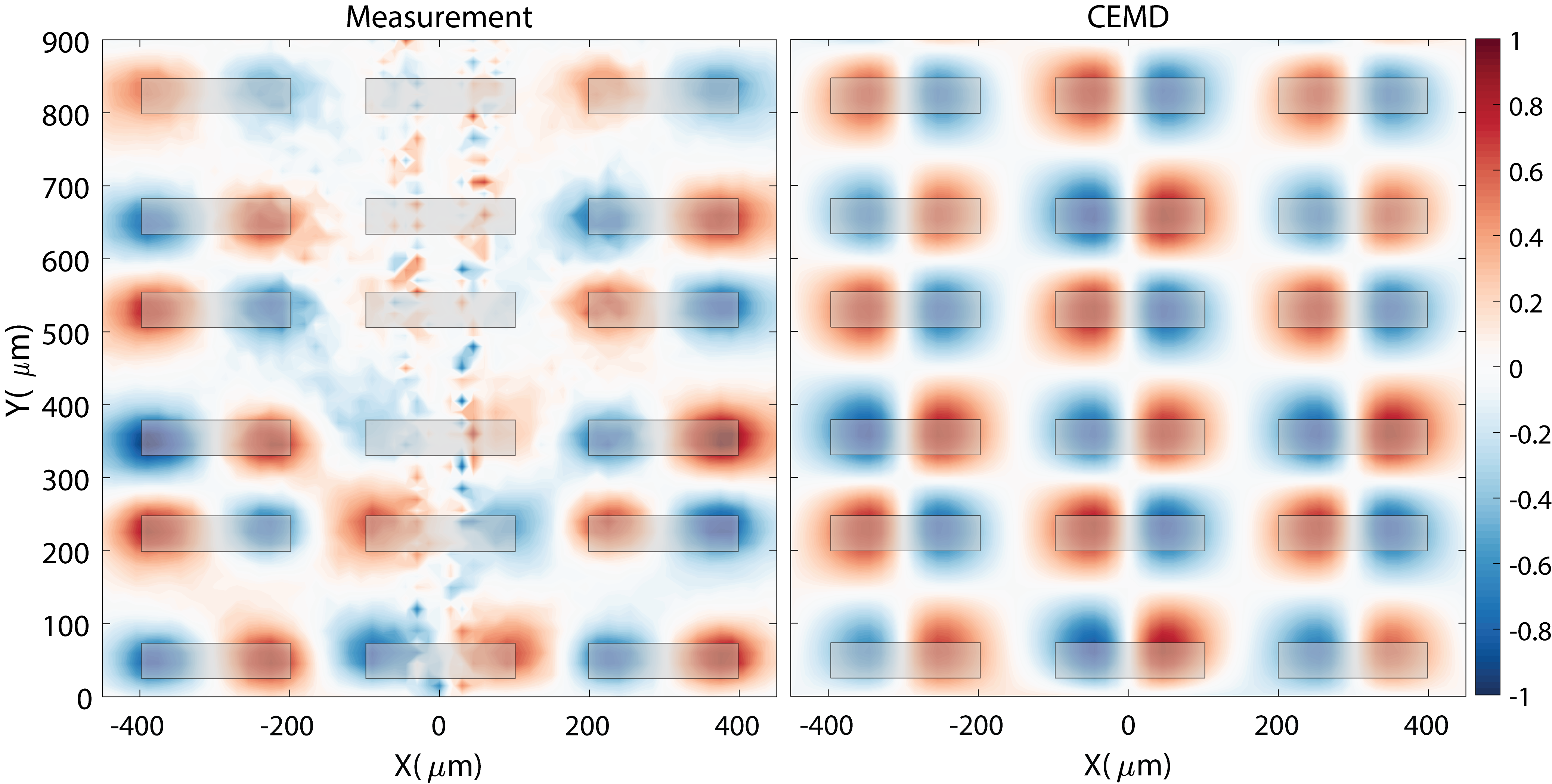}
\caption{Normalized $Im({E_z})$ electric field amplitude map over several unit cells of the array. (a) Measurement of the BIC at 0.39 THz, and (b) CEMD calculation at 0.381 THz. Both data sets have been normalized to their respective maximum field amplitudes. The excitation dipole is located at [X,Y]=[0 $\mathrm{\mu}$m, -70 $\mathrm{\mu}$m].  The gold rods are highlighted by the transparent grey areas.}
\label{fig:5_field_map}
\end{figure*}

The spatial extent of the field in the direction perpendicular to the long axis of the rods illustrates the formation of the BIC, as it is shown in Fig.~\ref{fig:2_illustration}(b). In this figure, the field emitted by \textit{Tx} locally couples into the BIC on the rod where the emitter is positioned. The out-of-plane field component ($E_z$) associated to this BIC is then measured as \textit{Rx} is moved away from \textit{Tx}. The lateral distance to the emitter increases successively from bottom to top as also graphically indicated by the rods on the left side of Fig.~\ref{fig:2_illustration}(b). In these measurements, we can see the time evolution of the pulse over the surface, showing pronounced oscillations of the near-field close to the rods. These long-lived oscillations can be associated to the symmetry protected BIC on the surface of the array. As discussed above, this BIC emerges from the resonant response of the individual rods, which are anti-symmetrically coupled with each other through a SLR in the array. Therefore, it is not surprising that the near-field amplitudes are maximum at the positions of the rods. The out-of-phase oscillation of adjacent rods in the unit cell of the array is visible in the near-field measurement. This out-of-phase oscillation is highlighted in Fig.~\ref{fig:2_illustration}(c), where the field amplitudes measured in the two rods of a dimer, indicated in Fig.~\ref{fig:2_illustration}(b) with the dashed lines, are plotted as a function of time. We note that the near-field oscillation decays as the EM energy spreads over the full surface of the sample, and not due to radiation losses. The unavoidable finite size of the array will finally lead to radiation leakage at the edges.

Figure~\ref{fig:3_spectra} displays the spectra obtained by Fourier transforming the THz transients measured at different positions in Fig.~\ref{fig:2_illustration}(b). These spectra show a narrow peak at 0.39 THz, in excellent agreement with the expected frequency of the BIC obtained from the CEMD calculation (Fig.~\ref{fig:1_2}). The line shape of the BIC, shown at the bottom of Fig.~\ref{fig:3_spectra} for two positions at rods in the same unit cell, is slightly asymmetric. This asymmetry can be explained with the calculated dispersion displayed in Fig.~\ref{fig:1_2}: The tail in the spectrum of the BIC at lower frequencies stems from the fact that the associated resonance is excited at non-zero in-plane wave vectors that are accessible for the dipole source, and to the non-negligible radiation leakage at these frequencies. In contrast, the near-field intensity peak decays abruptly at higher frequencies due to the absence of a resonance at those frequencies (see the dispersion plot in Fig.~\ref{fig:1_2}). Note also that, as shown in the bottom plot of Fig.~\ref{fig:3_spectra}, the spectra measured in adjacent rods have  the same resonant frequency and similar amplitude, differing only in the phase.

\begin{figure*}
\centering
\includegraphics[width=0.90\textwidth]{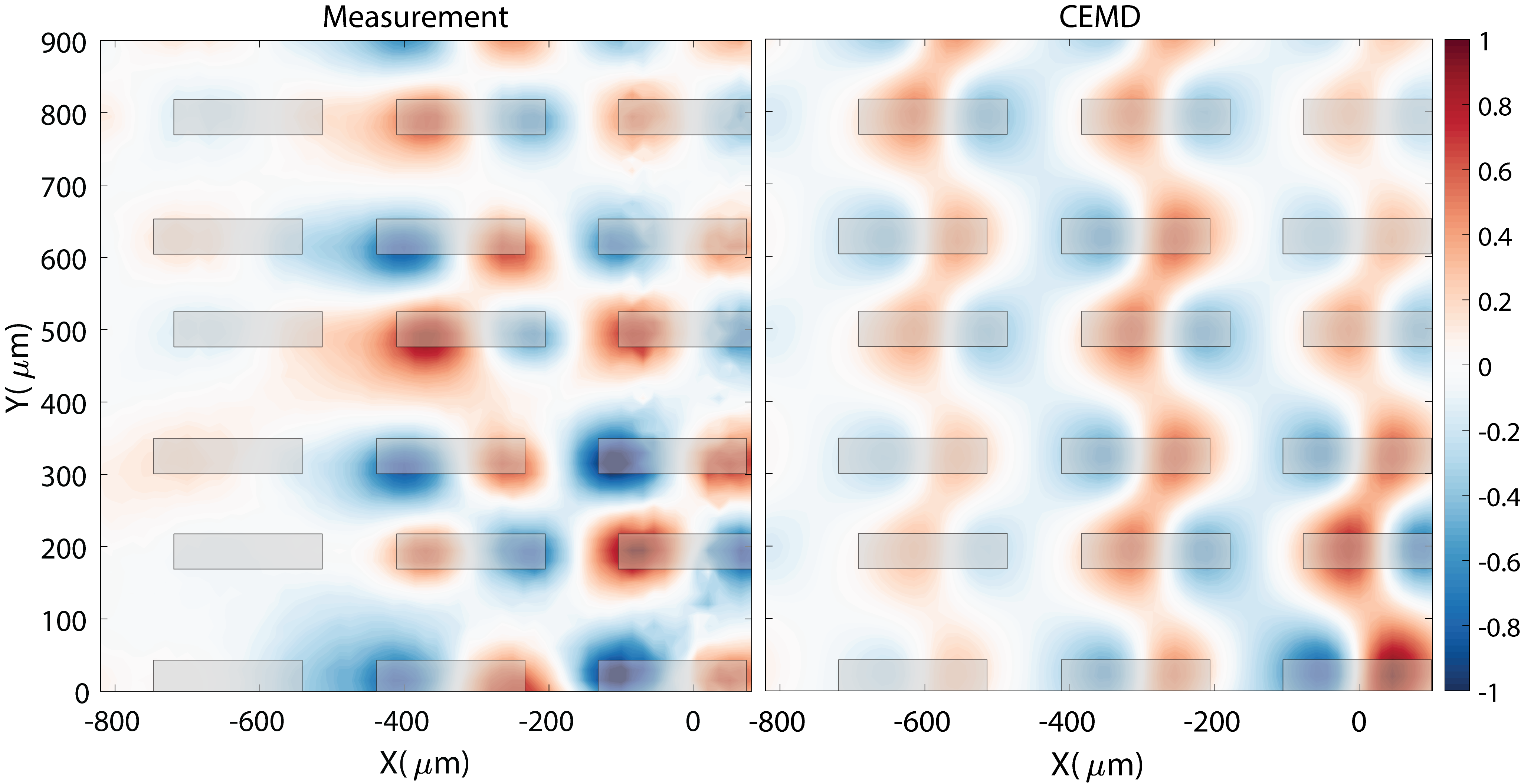}
\caption{Normalised $Im({E_z})$ electric field amplitude plotted at 0.39 THz over several unit cells, with the measurement of the BIC shown on the left and a calculation at 0.381 THz based on the CEMD theory, shown on the right. Both data sets have been normalized to their respective maximum field amplitudes. The excitation dipole is located at [X,Y]=[\SI{0}{\micro\meter},\SI{-70}{\micro\meter}]. The location of the gold particles are highlighted by the transparent grey areas.}
\label{fig:6_field_map_disp}
\end{figure*}

A map of the field-profile distribution of the BIC can be obtained by recording THz transients along a raster scan, Fourier transforming the transients, and plotting the THz amplitude at the frequency of the BIC. The results of these measurements and analysis are shown in Fig.~\ref{fig:5_field_map}(a). This iso-frequency amplitude map has been made by measuring 80 ps long transients at spatial intervals of 15 $\mathrm{\mu}$m, which are Fourier transformed and normalized to the global maximum to obtain the normalized field amplitude. Figure~\ref{fig:5_field_map}(a) shows the imaginary component of $E_z$ at 0.39 THz extending over the array, with the measurements on the left panel and the CEMD calculations on the right panel~\cite{Abujetas2021}. Note that the calculations have been done without any adjustable parameter. Despite some artifacts in the measurements around the $x = 0$ line, resulting from the laser beam intensity scattered on the emitter and reaching the detector, the calculations nicely reproduce the experimental BIC local field. Both theory and experiment show the phase shift across successive rods in the same unit cell, beating out-of-phase from each other. This field amplitude distribution corroborates the argument that a pair of parallel out-of-phase resonant dipoles per unit cell are responsible for the symmetry-protection of the BIC~\cite{Abujetas2019c}. An iso-frequency amplitude map measurement at a slightly different frequency (0.42 THz) is shown in the supplemental information (Fig.~\ref{fig:S3}). As appreciated in this measurement, the localized emission behavior of the dipole is recovered at other frequencies than the BIC frequency, where the radiation losses are significant and the extension of the field into the array is significantly less compared to the BIC.

So far, research on symmetry protected BICs has mainly focused on dielectric structures that rely on mirror and in-plane rotation symmetry protection to stabilize the BIC. These systems have been investigated by breaking the symmetry to form quasi-BICs that can be coupled to the continuum~\cite{Koshelev2018b}. However, the mirror symmetry can be relaxed to only include $\pi$-rotation symmetry, which in the case of single in-plane dipolar particles (e.g. metallic rods) will lead to an exceptional robustness of the BIC to relative displacements of the rods in the dimer. We have tested this remarkable property on an array of dimer rods in which the rods are displaced by 30 $\mathrm{\mu}$m in the $x$-direction, as shown in the measurements of Fig.~\ref{fig:6_field_map_disp}(a)  and the CEMD calculations of (b). The mirror symmetry of the dimer is broken in this sample, but the in-plane $\pi$-rotation symmetry is preserved. Both experimental and theoretical near-field maps shown in Fig.~\ref{fig:6_field_map_disp} exhibit an intricate combination of amplitudes that maintain the out-of-phase condition of each (dipole) rod within the unit cell throughout the metasurface. We have also measured the transmittance through this sample (Fig.~\ref{fig:S5} in the SI), where only the broad super-radiant mode ($\Lambda^{+}$) is observed, fully supporting the robustness of the BIC symmetry protection mechanism. 

\section{Conclusion}
With a double THz near-field technique, that allows the local excitation and near-field detection of broadband THz pulses, we have been able to map for the first time the electromagnetic field of symmetry protected BICs over extended areas. This investigation has been done in a periodic array of dimer rods supporting surface lattice resonances with a superradiant and a subradiant character. The near-field mapping of the BICs, associated to the subradiant mode, allows to unveil the field-symmetry protection that suppresses far-field radiation. This suppression of the radiation leakage leads to modes with exceptionally long (diverging) life times and high Q-factors that cannot be detected in the far-field. By displacing the rods, we demonstrate that mirror symmetry is not necessary for the formation of BICs and only $\pi$-rotation symmetry is required. This property, confirmed by theoretical calculations based on a coupled electric-magnetic dipole theory in the frequency domain, makes symmetry protected BICs exceptionally robust to relative displacements of the rods, which makes the fabrication of open resonant cavities with arbitrarily high Q-factors deemed simple. Although this investigation has been done at THz frequencies, it can be extended to other frequencies by scaling the dimensions and using non-absorbing materials.

\section*{Funding}
Nederlandse Organisatie voor Wetenschappelijk Onderzoek (NWO) (Vici 680-47-628); Spanish Ministerio de Ciencia e Innovación (MICIU/AEI/FEDER, UE) through the grants MELODIA (PGC2018-095777-B-C21) and NANOTOPO (FIS2017-91413-EXP), and Ministerio de Educación, Cultura y Deporte through a PhD Fellowship (FPU15/03566). 

\section*{Acknowledgements}
We acknowledge Koen de Mare for help in the development of the setup.

\bibliography{references}

\onecolumngrid
\clearpage

\section*{Supplementary Materials}

\vspace{5mm}

\textbf{Contents:}

\vspace{5mm}

 \begin{itemize}
 \item \textbf{S1 - Sample fabrication}
 \item \textbf{S2 - Coupled electric and magnetic dipole model}
 \item \textbf{S3 - Angle dependent transmission}
 \item \textbf{S4 - Near-field maps of THz pulse propagation in free space}
 \item \textbf{S5 - Iso-Frequency map away from BIC}
 \item \textbf{S6 - Surface lattice resonances versus BICs}
 \item \textbf{S7 - Robustness of the BICs to displacement}
 \end{itemize}

\vspace{5mm}

\section*{S1 - Sample fabrication}

The following recipe was used for the fabrication of the samples: First the fused silica substrate was cleaned thoroughly via immersion in an acetone bath for 5 minutes, followed by isopropyl alcohol (IPA) bath for 5 minutes and rinsed with dH2O. After the cleaning step, the surface of the substrate was activated with an Ion Wave Stripper using oxygen plasma at 600 W for 5 minutes, followed by a 2 minutes rinse in dH2O. Then, the substrate was vapour-coated with HDMS, to further improve the photoresist adhesion. The negative photoresist MaN-440 was spin-coated on the substrate for 30 seconds at 3000 rpm speed with an acceleration of 1000 rpm/s. It was then soft baked on the hot plate for 2 minutes at 100 $^{\circ}$C. The final photoresist film thickness was around 2 $\mu$m. The substrate was exposed using a Karl Suss MA-6 optical lithography at 365 nm for 100 seconds. This process was repeated 3 times with 10 seconds pause in between exposures. The contact mode between the mask and the substrate was set to ’hard’. This contact settings give a resolution of the structures greater than 1.5 µm. The exposed sample was developed for 90 seconds in MaD-532s developer and then rinsed for 2 minutes in dH2O. 2 nm of Titanium (Ti) and 100 nm Gold (Au) were deposited using BVR2008FC Electron Beam Evaporator. The evaporation rate for Ti and Au were 0.5 nm/s and 1 nm/s, respectively. The lift off was performed in acetone. Finally, the sample was rinsed for 2 minutes in dH2O.


\section*{S2 - Coupled electric and magnetic dipole model}

The transmittance through a rod dimer metasurface (Fig.~\ref{fig:1_2}) as a function of frequency and angle of incidence was calculated quasi-analytically through our coupled electric and magnetic dipole theory for an infinite planar array~\cite{Bulgakov2014,Abujetas2018a}, extracting the Q-factors of the corresponding resonances from the resulting spectra. The response of the metal rods is accounted for from their in-pane polarizabilities along the rod axis, which were numerically calculated through the free software SCUFF based on the method of moments~\cite{Abujetas2020a,SCUFF1}, using the SCUFF-scattering code. Near-field maps shown in Figs.~\ref{fig:5_field_map} and~\ref{fig:6_field_map_disp} were also computed by means of our CEMD formalism; nonetheless, rather than a plane wave scattering problem, a spatial Green function calculation is carried out~\cite{Abujetas2021}, placing the dipole source at one of the (fixed) arguments and scanning the other one on a plane above the metasurface.

\clearpage

\section*{S3 - Angle dependent transmission}

Far-field THz transmittance measurements were done as a function of the angle of incidence in a broad range of angles. These measurements are shown in Fig.~\ref{fig:S1}, where are (a) is the transmitted THz transient and (b) represents the THz transmittance spectra. We note the short transients and concomitant broad spectra for all the angles, in contrast with the narrow resonance calculated using the CEMD model and shown in Fig.~\ref{fig:1_2} of the manuscript. The reason for the non-observation of the narrow resonance is the finite spectral resolution of the THz measurements, which is limited by reflections in the sample substrate, making impossible to resolve these ultra-narrow resonances. The calculations of the Q-factor displayed in Fig.~\ref{fig:1_2}(c) of the manuscript show a maximum linewidth of 4 MHz at 10 deg, while the maximum frequency resolution that we can achieve using THz-TDS is 7 GHz. The broad resonance in Fig.~\ref{fig:S1}(b) is described by the imaginary component of the eigenvalue given by Eq.~\ref{eq:2}, and associated to the symmetric (super-radiant) SLR. 

\setcounter{figure}{0}
\renewcommand\thefigure{SI.\arabic{figure}} 
\begin{figure}[h!]
\centering
\includegraphics[width=0.90\textwidth]{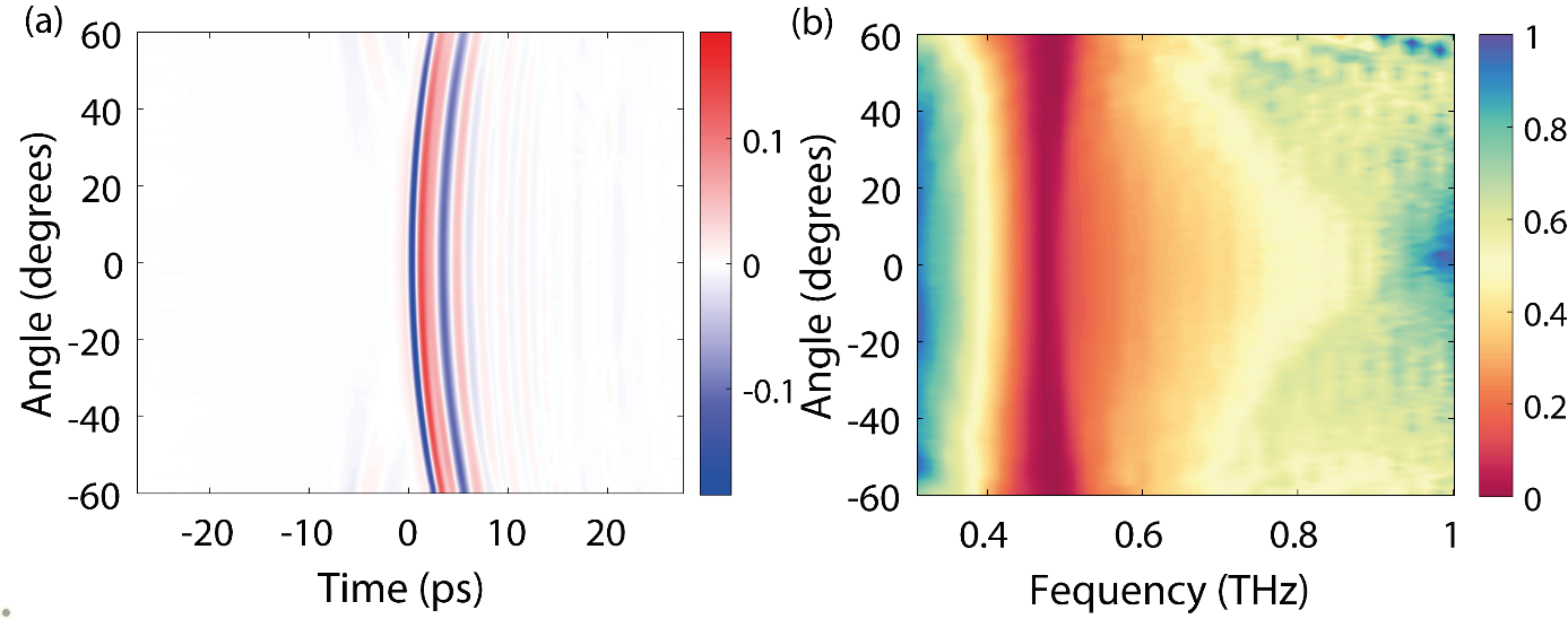}
\caption{(a) THz transients and (b) spectra of the transmission through the 2D periodic lattice of gold dimer rods measured at different angles.
}
\label{fig:S1}
\end{figure}

\clearpage

\section*{S4 - Near-field maps of THz pulse propagation in free space}

\begin{figure}[h!]
\centering
\includegraphics[width=0.70\textwidth]{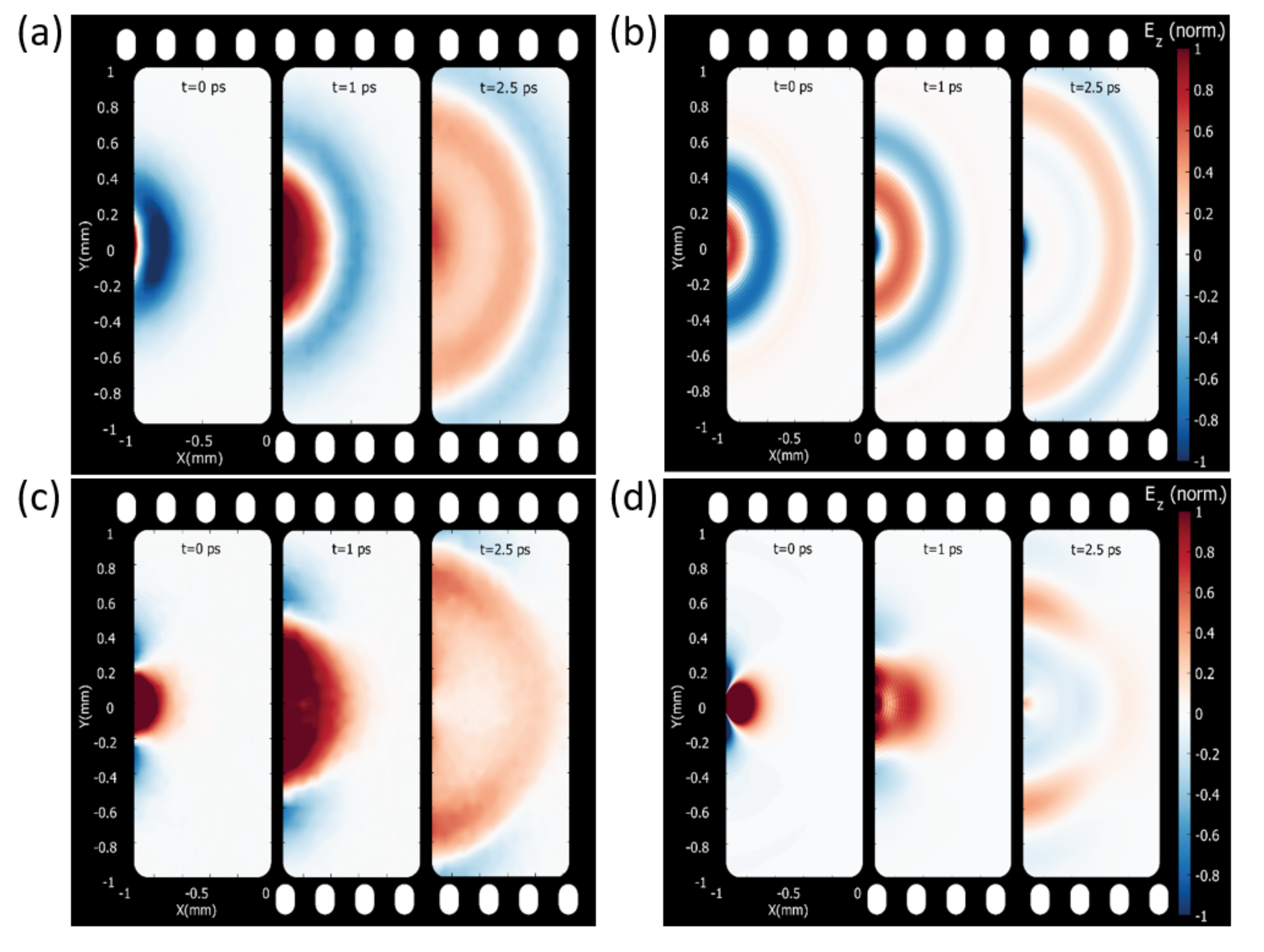}
\caption{Comparisons between measured and simulated fields propagating from a dipole source. (a) Measured THz electric field component parallel to the transition dipole moment in the perpendicular plane, i.e., $E_z$-field component in the $x$-$y$ plane of a THz near-field photoconductive antenna with a 5 $\mu$m gap between electrodes oriented in the z-direction. (b) Calculated $E_z$-field component in the $x$-$y$ plane of a point dipole with dipole moment along $z$. (c) Measured THz electric field component parallel to the transition dipole moment in the parallel plane, i.e., $E_x$-field component in the $x$-$y$ plane of a THz near-field photoconductive antenna oriented in the $x$-direction . (d) Calculated $Ex$-field component in the $x$-$y$ plane of a point dipole with dipole moment along $x$.
}
\label{fig:S2}
\end{figure}

The double THz near-field microscope was calibrated and tested by measuring the THz field amplitude of a pulse emitted by one of the probes used as an emitter. Three time snapshots of this measurement are shown in Fig.~\ref{fig:S2}(a) and Fig.~\ref{fig:S2}(c). These measurements correspond to the amplitude at different times for the field component parallel to the dipole moment of the emitter, and measured in the perpendicular (Fig.~\ref{fig:S2}(a)) and parallel planes Fig.~\ref{fig:S2}(c)). The measured fields resemble the emission of a radiating dipole, as can be compared with the propagation calculations shown in Figs.~\ref{fig:S2}(b) and~\ref{fig:S2}(d). The calculations are based on the dipole model as can be found in textbooks, in this case taken from ’Principles of Nano- optics’ by Novotny and Hecht (2006). The dipole is modeled by having a small charge displacement $\mathrm{d}s(t)$ oscillating in time and expressed using the retarded time $t_r=t-c/r$, which causes a time dependent dipole moment, equal to $p(t)=q \cdot \mathrm{d}s(t)\cdot \hat{z}$.  Using the retarded potential formulation, the electric field is calculated with:
\setcounter{equation}{0}
\renewcommand{\theequation}{SI.\arabic{equation}}
\begin{equation}
    E = \dfrac{1}{4\pi\epsilon_0}\left[ \dfrac{1}{c^2r}\dfrac{\partial^2}{\partial t^2} \left(\hat{r}\times p(t_r) \right)\times\hat{r} + \left(\dfrac{1}{r^3} + \dfrac{1}{c^2 r} \dfrac{\partial}{\partial t} \right) \left(3\hat{r}\left[ \hat{r}\cdot p(t_r) \right] - p(t_r) \right)  \right].
\end{equation}
The full movies of the dipole field emission are also available as supplemental material.
These measurements and calculations illustrate that our near-field microscope can be used for the local excitation and the spatially resolved detection of the THz electric field amplitude as a function of time.

\clearpage

\section*{S5 - Iso-Frequency map away from BIC}

\begin{figure}[h!]
\centering
\includegraphics[width=0.85\textwidth]{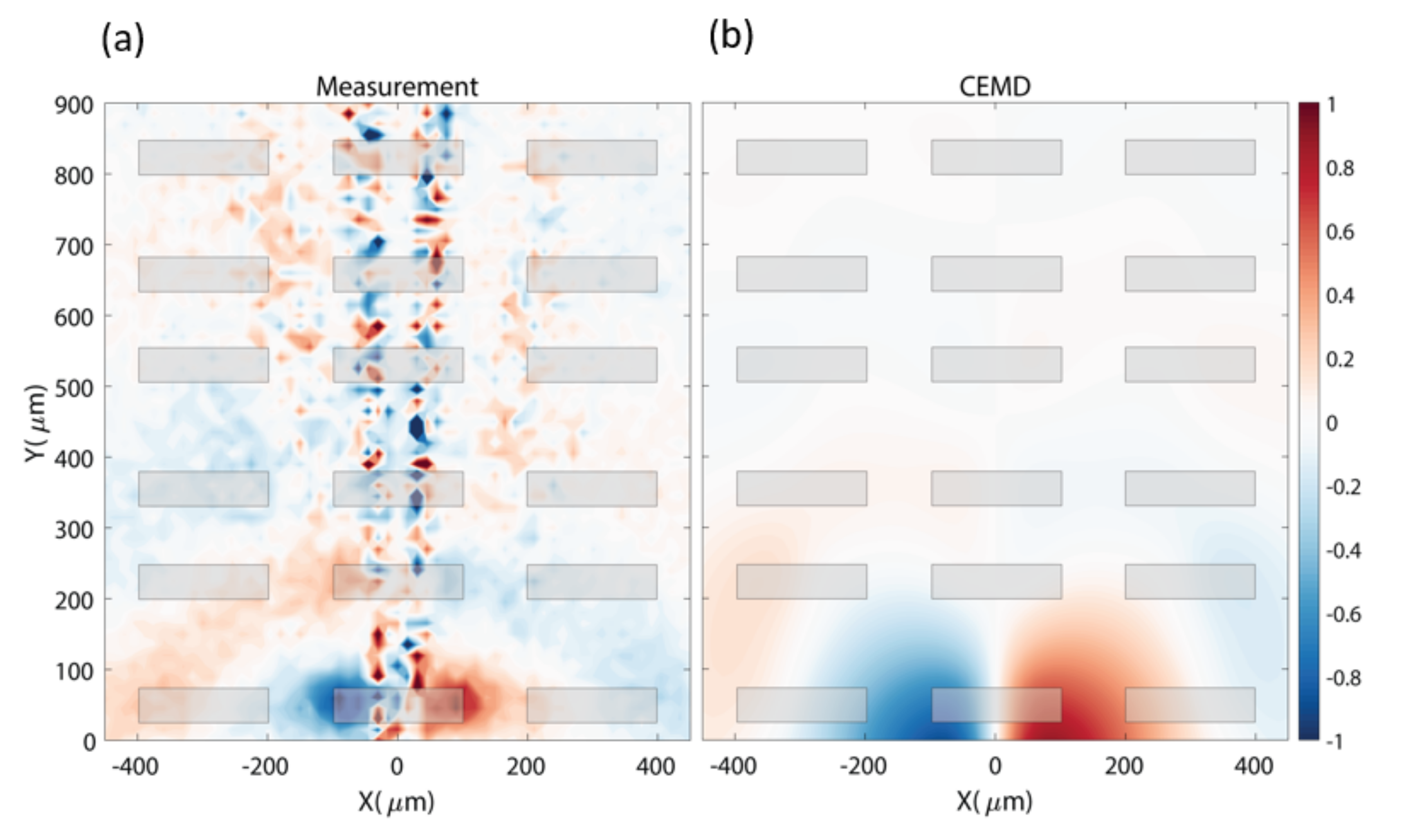}
\caption{Normalized $Im({E_z})$ electric field amplitude maps measured at 0.42 THz over several unit cells. (a) Measurements and (b) calculations based on the CEMD model. Both data sets have been normalized to their respective maximum field amplitude. The excitation dipole is located at [X,Y]=[0 $\mu$m, 70 $\mu$m]. The gold particles are highlighted by transparent grey areas.
}
\label{fig:S3}
\end{figure}

The imaginary component of the $E_z$-field amplitude measured at 0.42 THz is shown in Fig.~\ref{fig:S3}(a). This field amplitude in mainly localized near the emitter, extending much less into the array than at the frequency of the BIC (Fig.~\ref{fig:5_field_map} of the manuscript), as it is expected if radiation losses are significant. The measurement is in agreement with the CEMD calculations shown in Fig.~\ref{fig:S3}(b), where the equivalent near-field amplitude maps are extracted from the time-harmonic Green functions at a similar frequency, with one of the arguments determining the dipole source whereas the other scans a horizontal plane located at the height of the detector’s position. The only difference between measurements and calculations is the dipole source height, which is located higher above the metasurface while maintaining a sub-wavelength distance to the metasurface, to avoid the large fields arising near the electric dipoles representing the rods in this model. The artifacts in the measurements along the y-direction above the THz emitter are due to the scattered laser light on the emitter (used for the generation on THz radiation) reaching the THz detector.

\clearpage

\section*{S6 - Surface lattice resonances versus BICs}

\begin{figure}[h!]
\centering
\includegraphics[width=0.65\textwidth]{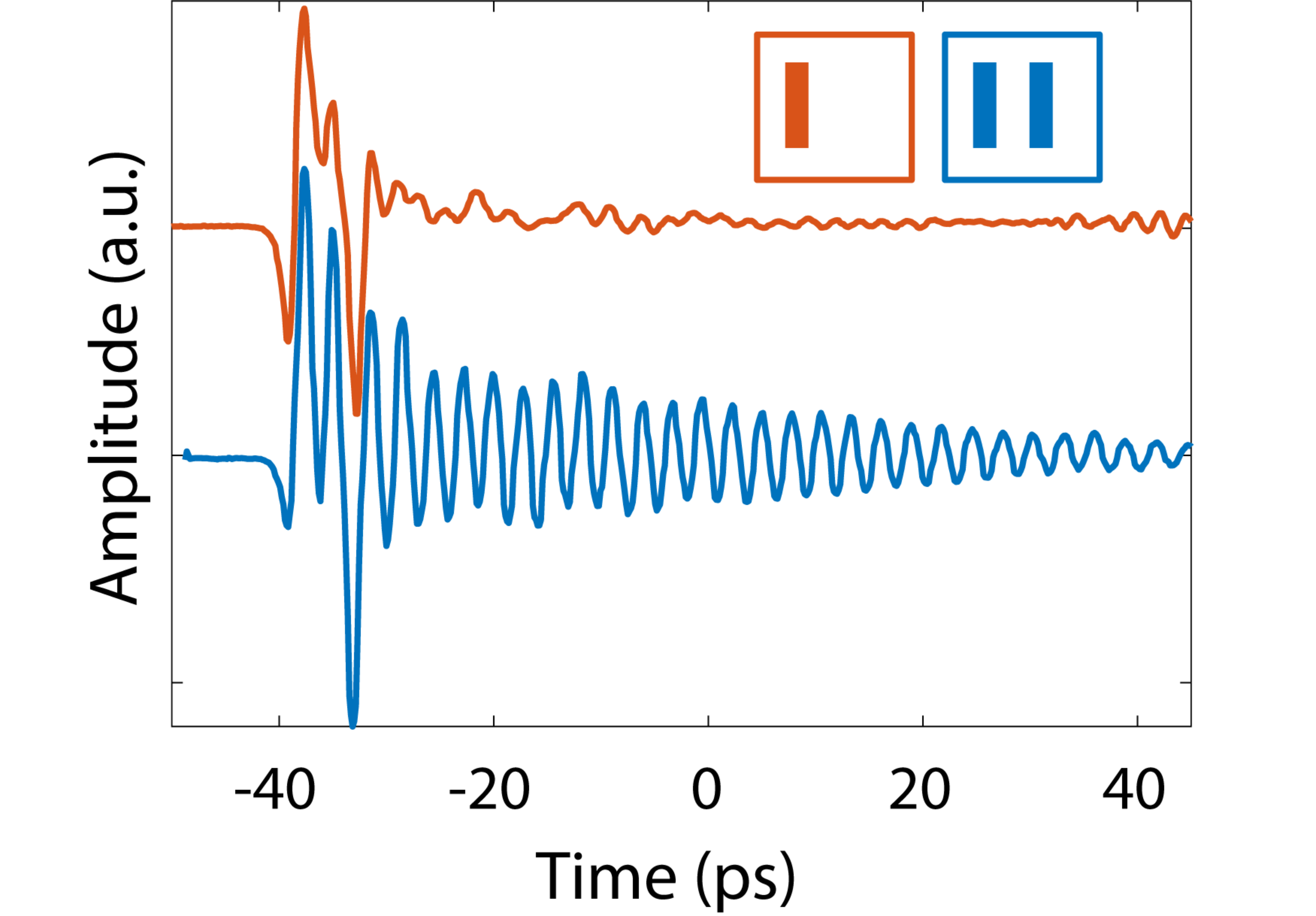}
\caption{Transients of the electric field $E_z$ measured on a rod located two unit cells away from the source. Blue curve: measurement on the array discussed in the manuscript. Red curve: measurement on a similar array with only one rod per unit cell (see also inset).
}
\label{fig:S4}
\end{figure}

To validate that the measured near-fields in $\pi$-rotational symmetry arrays at the BIC frequency are indeed due to BICs and not just to surface lattice resonances propagating on the sample, we have recorded the out-of-plane electric field $E_z$ on a rod located two unit cells away from the emitter for two different arrays. The first array corresponds to the one investigated in the manuscript (Fig.~\ref{fig:1_1}), while the second is a similar array with only one rod per unit cell. The near-field of the first array corresponds to the blue curve in Fig.~\ref{fig:S4} and the near-field of the second array is the red curve. The field in this last array quickly decays due to out-coupling of the EM energy to the far-field, while the field associated to the BIC rings-down for more than 80 ps. We also note that the decay of the amplitude is mainly due to the distribution of the EM energy over the surface of the array, and not to radiation losses.

\clearpage

\section*{S7 - Robustness of the BICs to displacement}

\begin{figure}[h!]
\centering
\includegraphics[width=0.60\textwidth]{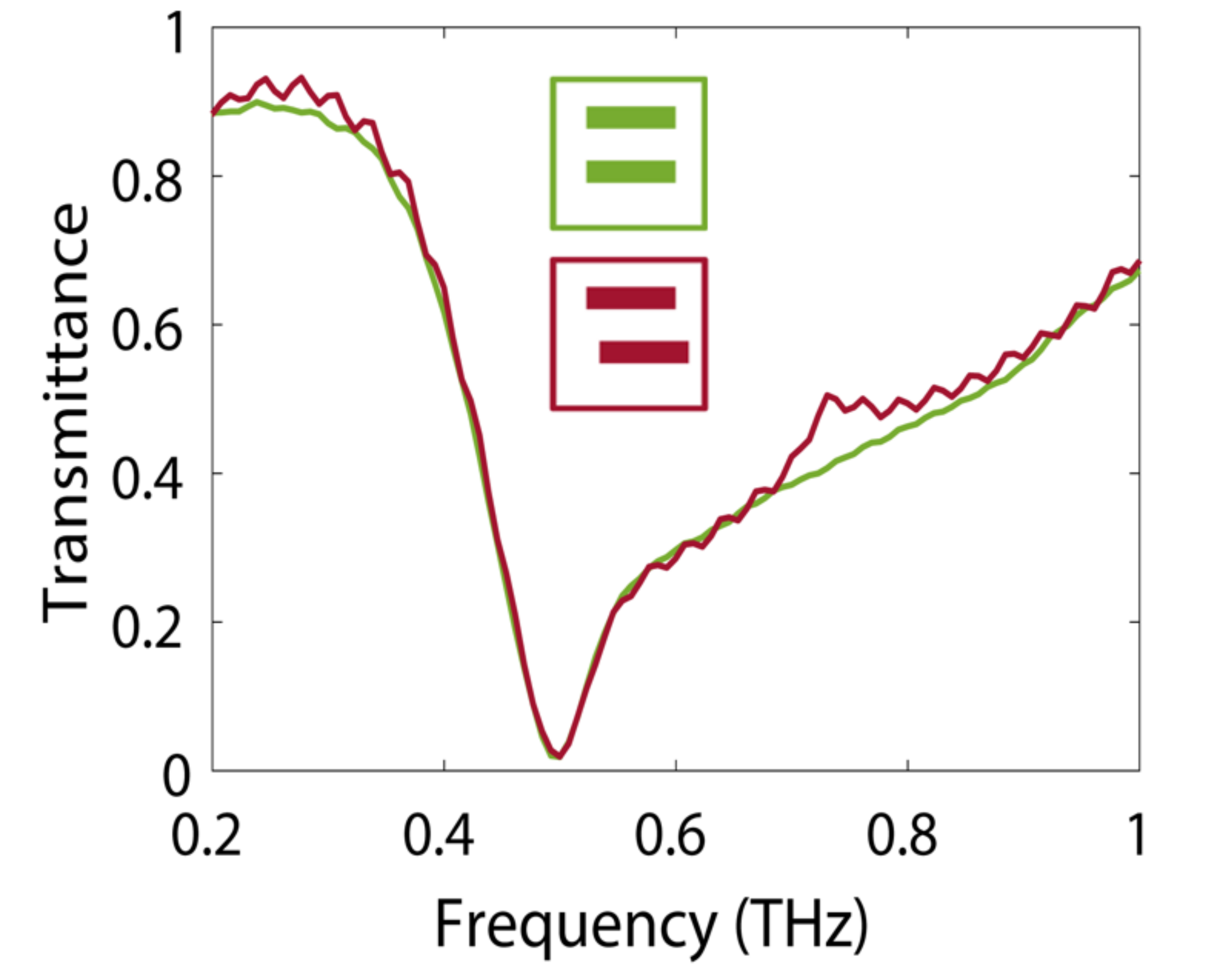}
\caption{Green curve: Transmittance measurement at normal incidence through the array with aligned dimer rods. Red curve: Transmittance through the array with the rods in the dimer displaced by 30 µm.
}
\label{fig:S5}
\end{figure}

The robustness of BICs in arrays of dipole dimers to their relative displacements, due to the preserved -rotational symmetry,  has been investigated by measuring a sample with the same lattice constant and length of the rods as the one described in the manuscript, but where the rods in the dimer are displaced by 30 $\mu$m in the $x$-direction, as depicted in the inset of Fig.~\ref{fig:S5} (near-field maps are shown in Fig.~\ref{fig:6_field_map_disp} of the manuscript). The green curve in this figure corresponds to the transmittance measurement of the sample with aligned rods, while the red curve corresponds to the transmittance of the sample with displaced rods. The absence of a resonance at  0.4 THz illustrates that the displacement of the rods does not lead to an increase of the radiation losses of the BIC. 

\end{document}